# Electron Beam Production by Pyroelectric Crystals

James D. Brownridge[a]
Department of Physics, Applied Physics, and Astronomy
State University of New York at Binghamton, P.O. Box 6016 Binghamton, New York 13902-6016

Stephen M. Shafroth[b]
Physics and Astronomy Department, University of North Carolina at Chapel Hill, Chapel Hill North Carolina 27599-3255

## Abstract

Pyroelectric crystals are used to produce self-focused electron beams with energies greater than 170 keV. No high voltage power supply or electron gun is needed. The system works by simply changing the temperature of a crystal of $LiNbO_3$ or $LiTaO_3$ by about 100°C in dilute gas. Electron beam energy spectra as well as positive-ion-beam energy spectra and profiles are shown. A change in the crystal temperature of 100°C will cause a spontaneous change in polarization. The change in polarization will be manifested by a change in charge on the surface of the crystal. It is this uncompensated charge that produces the electric field, which accelerates the electrons, or the positive ions and gives rise to the plasma, which in turn focuses them. The source of the accelerated electrons or positive ions is gas molecules ionized near the crystal surface. When the crystal surface is negative electrons are accelerated away from it and positive ions are attracted to the surface. These positive ions reduce the net negative charge on the surface thereby reducing the electric field, which causes the electron energy to decrease over time even though the focal properties remain unchanged. When the surface is positive the reverse obtains and the positive ion beam energy decreases over time as well. We will present video clips, photographic and electronic data that demonstrate many of the characteristics and applications of these electron beams.

## Introduction

Even though pyroelectric crystals have been known since ancient Greek times[1], surprising new effects and practical applications are constantly being discovered [1-9]. There is a voluminous literature on electron emission by pyroelectric crystals. Reference 10, a comprehensive review article and (11-15) are representative but none of them suggests that energetic electron beams are produced by such crystals. Much of the current industrial use of these crystals is in infrared detection, sensitive temperature change detectors and photonics, e.g., the detection of THz radiation[16] in $LiNbO_3$. For these reasons few studies of the behavior of these crystals in dilute gases have previously been done and so no one has previously discovered the phenomena being reported on here.

The first known naturally occurring pyroelectric crystals were kidney stones of lynx[1]. All pyroelectric crystals including artificially produced $LiNbO_3$ and $LaTaO_3$ are asymmetric crystals. They become electrically charged on heating and cooling and give rise to electric fields at their surfaces of up to $10^6$ V/cm[18]. The opposite side of a crystal which is heated from its + z base becomes positively charged on heating and negatively charged on cooling. This is due to the movement of Li+ and Nb- ions relative to the oxygen lattice[12]. If the crystal is immersed in a dilute gas the net charge on the crystal surface at any time will depend on the gas type and pressure as well because when the surface is negative, positive ions from the gas will bombard it and when the surface is positive, electrons from the ionized gas will also bombard it. Crystal x-ray emission occurs on heating and target x-ray emission on cooling when the -z base is exposed and light is produced both on heating and on cooling in the plasma adjacent to the crystal.

Chronologically, JDB[2] reported on the production of characteristic x-rays by cesium nitrate crystals in 1992. Then in 1999 he and S. Raboy wrote a comprehensive paper[3] describing the x-ray emission and light production in the plasma for several different crystals and gases. Also in 1999 Shafroth et. al. studied the time dependence of x-ray emission[4]. Further they studied the effect on total x-ray yield for five different gases and found a marked decrease in yield for $O_2$ which was due to the production of ozone in the plasma. Also in 1999 Shafroth and Brownridge reported on the use of pyroelectric crystals in the teaching of x-ray physics[5]. Then in 2001 JDB and collaborators reported on the production of energetic electrons by pyroelectric crystals in dilute gases[6]. They found that multiple, energetic (~100 keV) nearly monoenergetic electrons were produced on cooling these crystals when the - z base faced the detector. Later JDB and others[7] found that the maximum energy of the emitted electrons depended somewhat on gas type but strongly on pressure. Then JDB and SMS found[8] that if the crystals were ground to a cylindrical shape that focused beams of electrons were produced as the crystal cooled and focused beams of positive gas ions were produced as the crystal was heated. Most recently, S. Bedair et. al., using a LabVIEW program have studied the effect of pressure on the time dependent yield of crystal x-rays and target current[9].

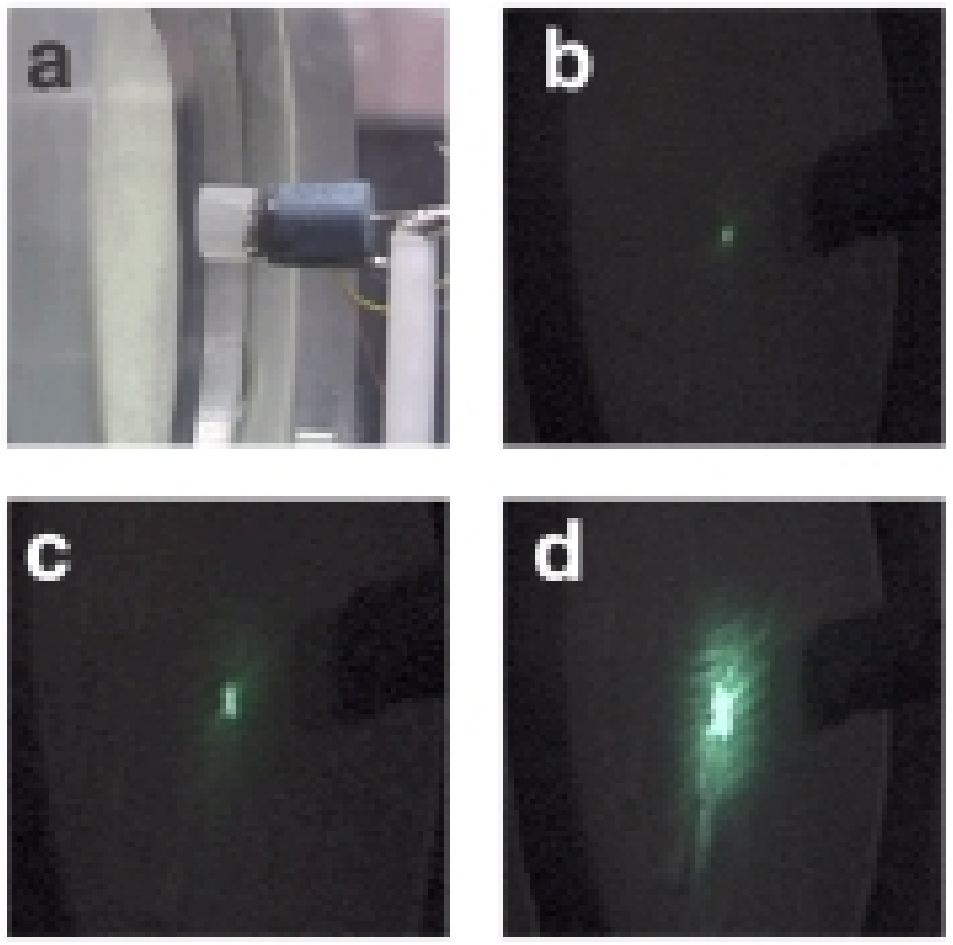

**Fig. 1.** (a) Photograph of the experimental arrangement with ZnS screen, crystal, and heater resistor, (b) Beam spot at ~0.5 mtorr, (c) ~3 mtorr, (d) ~8 mtorr, where the beam blows up. Note how the beam is defocused as pressure increases.

## Focused Electron Beams

This report describes the production of self-focusing electron beams arising from near pyroelectric crystal surfaces in dilute gases after the crystal has been heated and returned to room temperature. In one case a 5 mm diam x 5 mm cylindrical $LiNbO_3$ crystal in an atmosphere of <10 mtorr of dry $N_2$ was contained in a glass tube.

The crystal was heated to about 115 $^{0}$C and then allowed to cool to room temperature. Heat was supplied by passing ~100 mA through a wire-wound 60 ohm resistor which was epoxied to the crystal's + z base. After cooling to room temperature, spatially-stable electron beams were produced. They were made visible by placing a ZnS screen at the focal length (17 mm) of the crystal as illustrated in Fig. 1(a), which shows a photograph of the ZnS screen; the cylindrical crystal and the 60 ohm wire-wound resistor used to heat the crystal. We studied the effect of pressure on the beam. Typical results at ~0.5 mtorr 1(b), ~3 mtorr 1(c) and ~8 mtorr 1(d) are shown in Fig 1(b), 1(c), 1(d) respectively. As the pressure increased, the beam spot became more diffuse and brighter. At ~8 mtorr the beam blew up and its intensity dropped to zero. The dynamical behavior of the electron beam with pressure, which is illustrated in the video clip[17] suggests a gas-multiplication effect. The focusing effect may be due to plasma focusing. All results reported on in this paper were repeated many times.

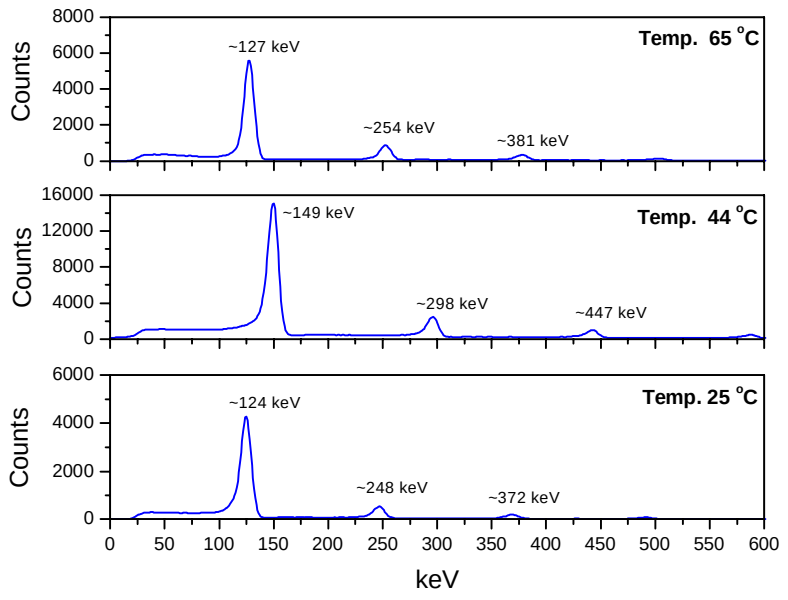

**Fig. 2.** Electron Energy Spectra taken with a 10 mm x 4mm dia $LiNbO_3$ crystal in 2.7 mTorr of $N_2$ .

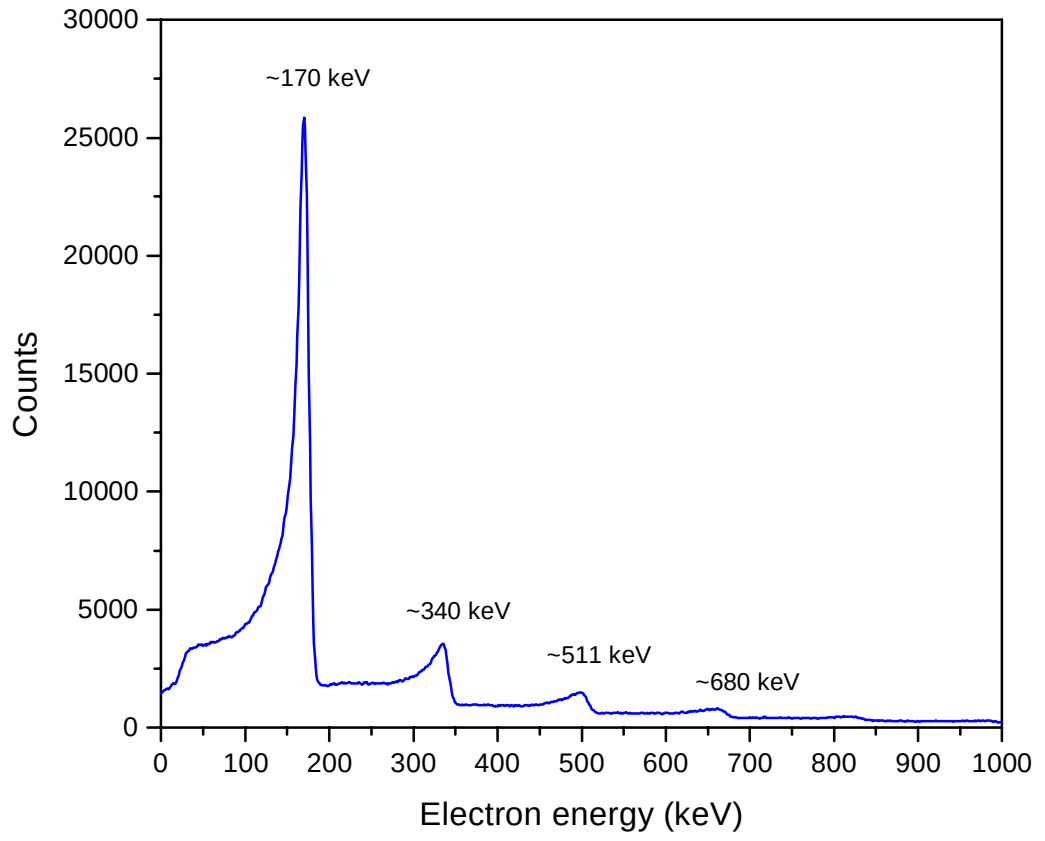

**Fig. 3.** Electron energy spectrum for highest energy electron beam yet obtained. Same conditions as Fig. 2.

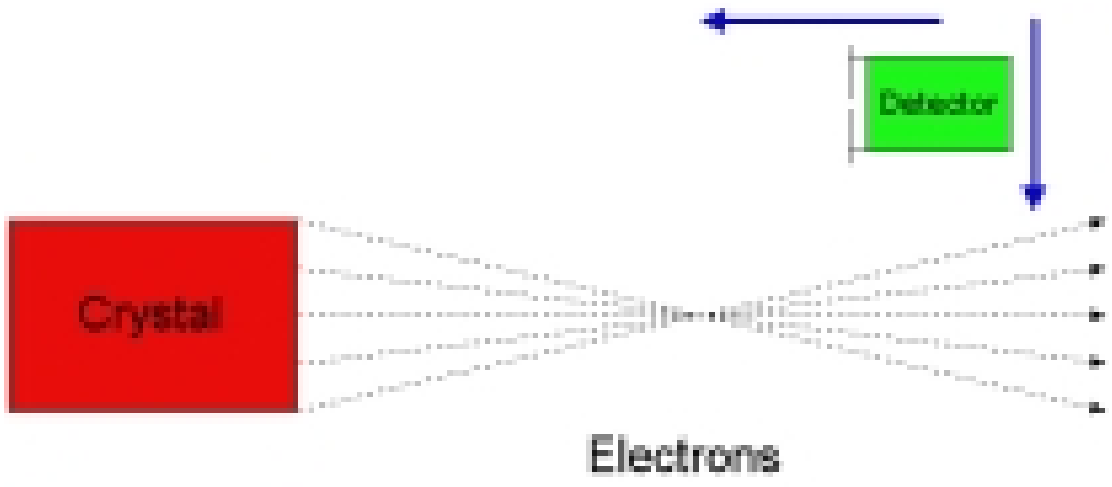

**Fig. 4**. Schematic diagram of the experimental arrangement for scanning the electron beam. The slit and detector move. The dotted lines show an idealized cartoon of the crystal focused electron beam.

Next a 100 μm surface-barrier electron detector was installed in place of the ZnS screen so that electron energy spectra could be recorded. The time-dependent electron-beam energy spectrum, which was taken in "snapshot mode", i.e. over an interval of <60 s, typically, exhibited discrete peaks at integral multiples of the lowest energy peak. This indicated that nearly monoenergetic multiple electron production[6] was occurring. Typical electron energy spectra are shown in Fig. 2. where the multiple peak effect due to pile up is shown at three different crystal temperatures. Fig. 3. shows the highest energy electron spectra so far obtained. The experimental arrangement for scanning the electron beam profile is shown schematically in Fig. 4. The crystal and heater assembly could be moved toward or away from the detector, which was useful for checking the focusing effect.

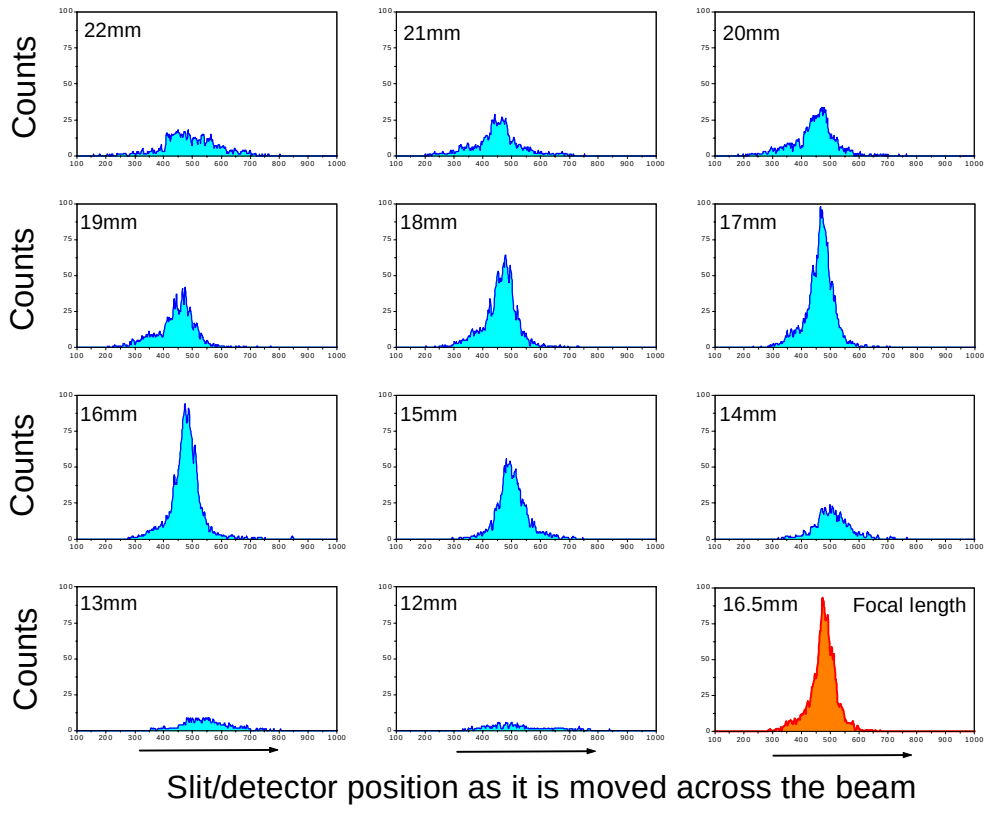

**Fig. 5**. Beam scans taken at different crystal-to-slit distances. Each scan lasted ~30 s. The focal length is ~16.5 mm. Profiles were taken in the order shown.

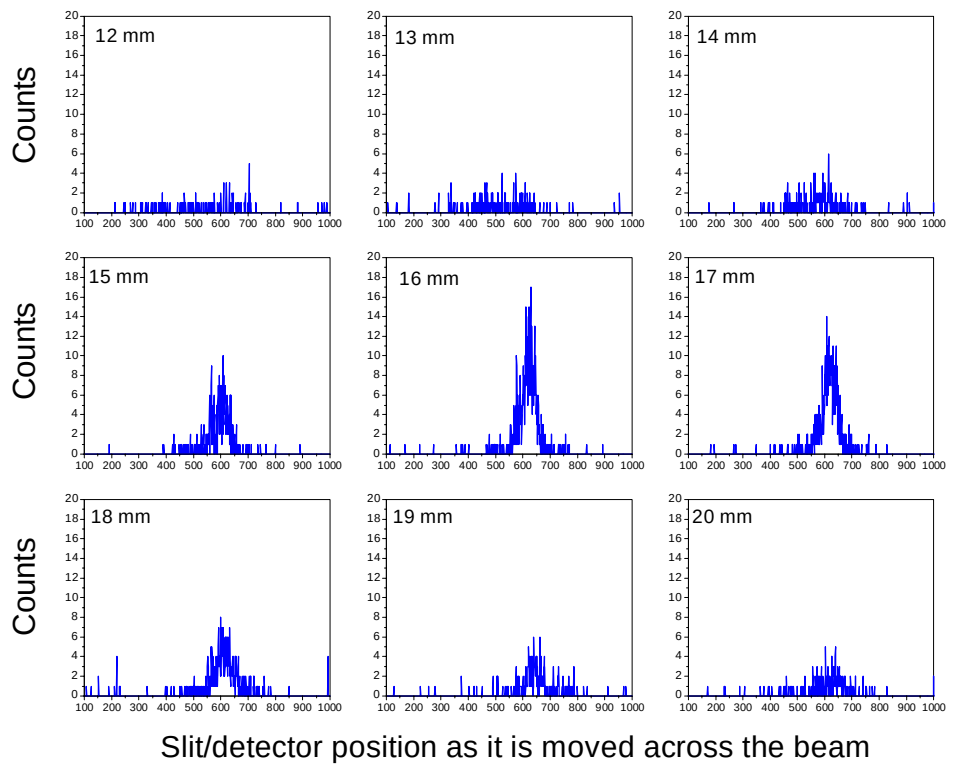

**Fig. 6.** Electron beam profiles taken 16 hours later. The focal length is still ~16.5 mm. The beam intensity is down after 16 hours of continuous electron emission, however it is still focused.

The detector was covered by a lead screen with a 0.1mm slit so that the beam profile could be obtained by moving the slit-detector assembly at constant speed across the beam at different crystal-to-slit distances. Fig. 5. shows results of ~ 30 s scans taken at different crystal-to-slit distances.

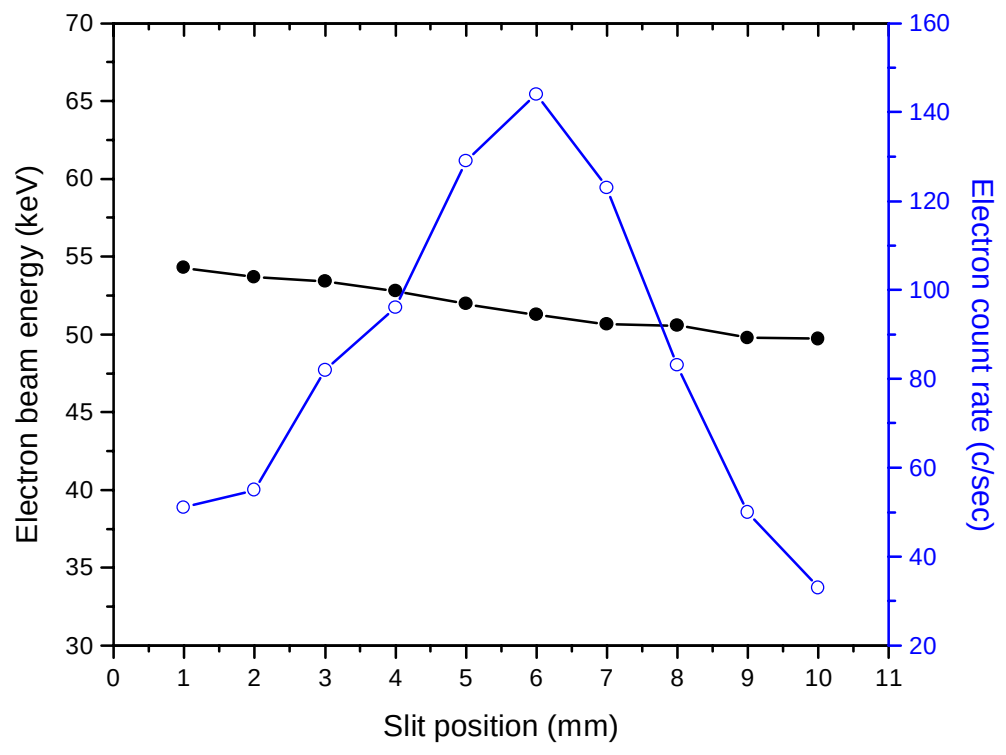

**Fig. 7**. The electron beam count rate profile for a 10 mm x 4 mm $LiNbO_3$ crystal (open circles). The electron beam energy (solid circles) is shown for each slit position. Note the linear decrease in energy as the scans progress and as time increases. (See Fig 8 for comparison).

All pulses corresponding to electron energies > 15 keV were counted for these scans. From this figure it can be seen that the best focus is obtained 16.5 mm crystal-to-slit distance. Fig. 6. shows electron beam scans taken 16 hours later. The focal length is still 16.5 mm. Fig. 7. obtained with a 4mm diam x 10 mm $LiNbO_3$ crystal shows the electron beam count rate profile as the detector and slit are scanned across the beam at a distance of 20 mm. The FWHM is about 4 mm. Also shown is the electron energy at each slit position. The uniform decrease in electron energy with slit position and time is similar to that described below.

## Electron Beam Energy Spectra

Once the crystal reached room temperature the beam energy decreased nearly exponentially with time (Fig. 8); faster with increasing pressure. In order to produce higher energy electrons the crystal was heated from the + z base to about 160 $^{0}$C and allowed to cool to room temperature at <10 mtorr.

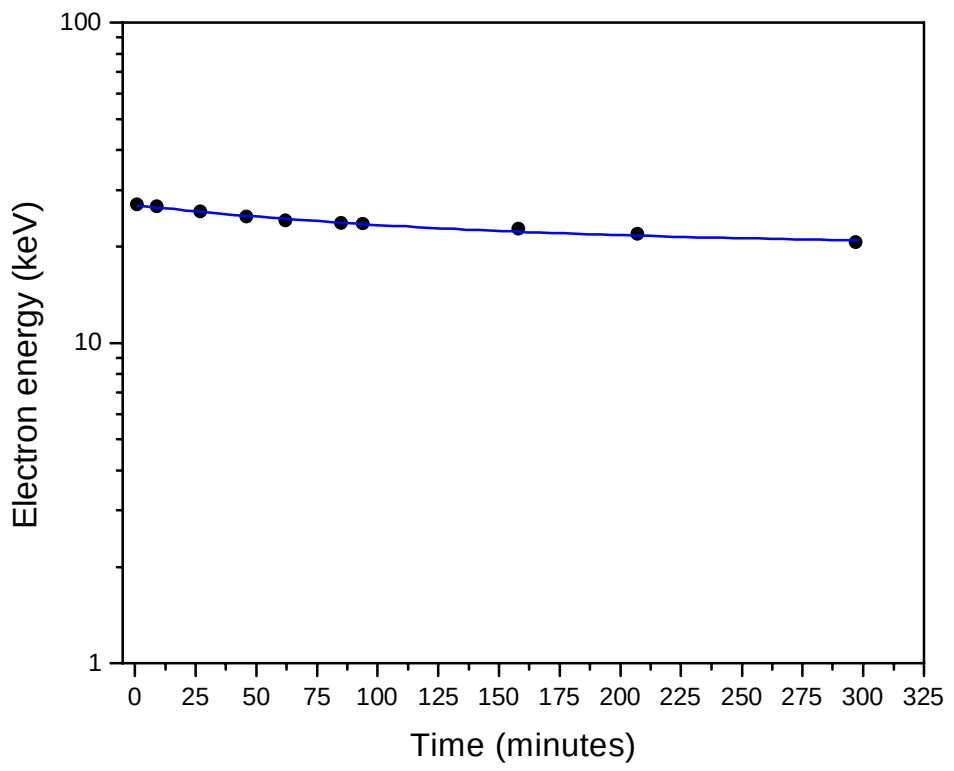

**Fig. 8.** Electron beam energy as a function of time after the crystal has reached room temperature. Note the nearly exponential decay of the beam energy.

The maximum energy electrons (170 keV) were produced after the crystal temperature had dropped to ~30 $^{0}$C. Independent proof that the electron beam energy was at least 80 keV (Au K edge) for a significant time was confirmed by irradiating gold and observing the K X-rays[2,3]. Gold K electrons were being ionized thereby producing Au K X-rays as long as the incident electron energy was greater than 80 keV i.e. starting as the crystal cools from 160 $^{0}$C when the increasing electron energy reached 80 keV and continuing until the maximum energy (170 keV) is reached and then until the electron energy falls to less than 80 keV, sometime after the crystal has reached room temperature.

## Phenomenological Description of These Processes

If the crystal is subjected to a rapid temperature change the polarization will be manifested briefly by the presence of an electric field whose strength is proportional to the surface charge density, which in turn is a function of the change in temperature and depends on the ambient pressure. When the crystal is in a reduced pressure environment and is subjected to the same temperature change, the polarization change is manifested by an electric field that lasts much longer; at a pressure of about $10^{-6}$ torr, it takes more than 30 hours for the resulting surface charge to be neutralized, whereas at 1 atm neutralization is virtually instantaneous[3]. The neutralization of the electric field produced by a change in polarization is due to bombardment of and attachment to the surface by positive ions. To summarize, as long as the net surface charge is sufficiently negative, electrons are accelerated away from the crystal in a focused beam whose energy can be up to 170 keV after which the energy of the electrons in the beam decreases nearly exponentially with time until the electrons become undetectable. It is remarkable that the beam becomes spatially stable once the crystal has reached room temperature, even before in some cases.

## Positive Ion Beams

We believe that singly charged positive ion beams arising from the residual gas have also been observed. For practical reasons they can best be studied by exposing the + z base of the crystal and taking measurements as the crystal cools, although the same phenomena can be observed on heating a crystal with its -z base exposed. Figure 9 shows an energy spectrum obtained over a period of 35 s with a crystal of $LiNbO_3$, 10 mm x 4mm dia heated to 170$^{0}$C after the crystal had cooled to ~35$^0$C . Here the gas was $N_2$ so the beam was presumably $N_2$+. The main component of the beam has an energy of ~100 keV but there are secondary components with energies of ~34 and ~62 keV, which we don't understand.

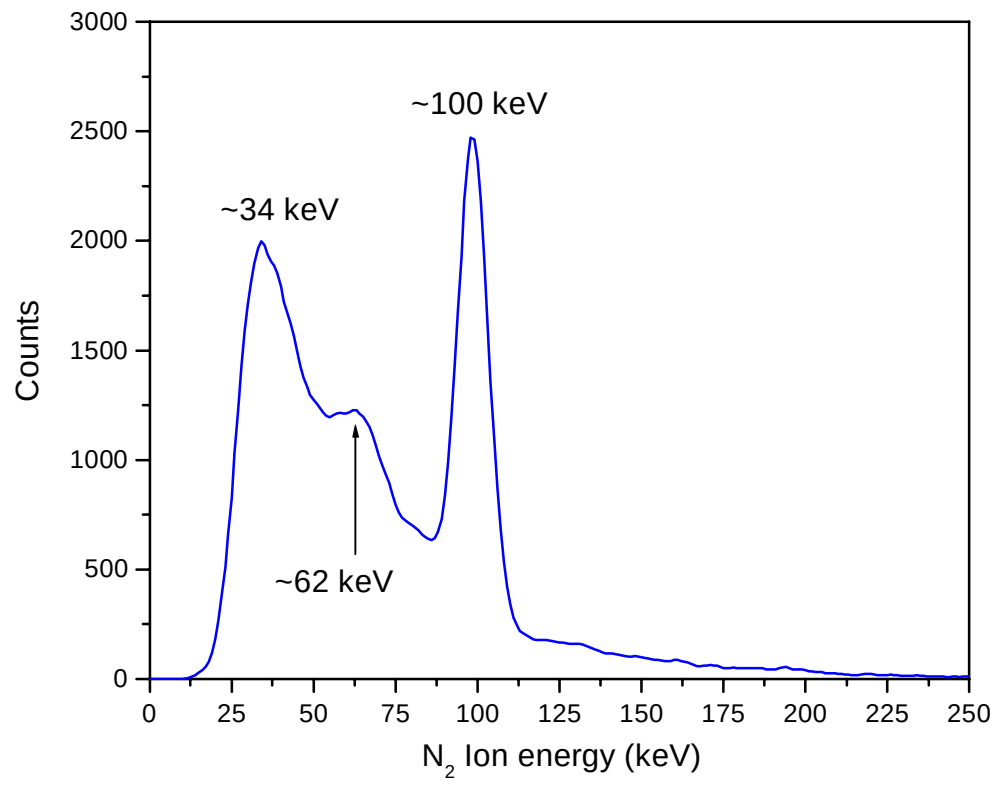

**Fig. 9.** Energy spectrum (30 s snapshot) of $N_2$+ ion beam obtained with a. $LiNbO_3$ crystal on cooling from 170 $^0$C. The + z base is exposed. The spectrum was obtained after cooling to 35 $^0$C. The gas was $N_2$.

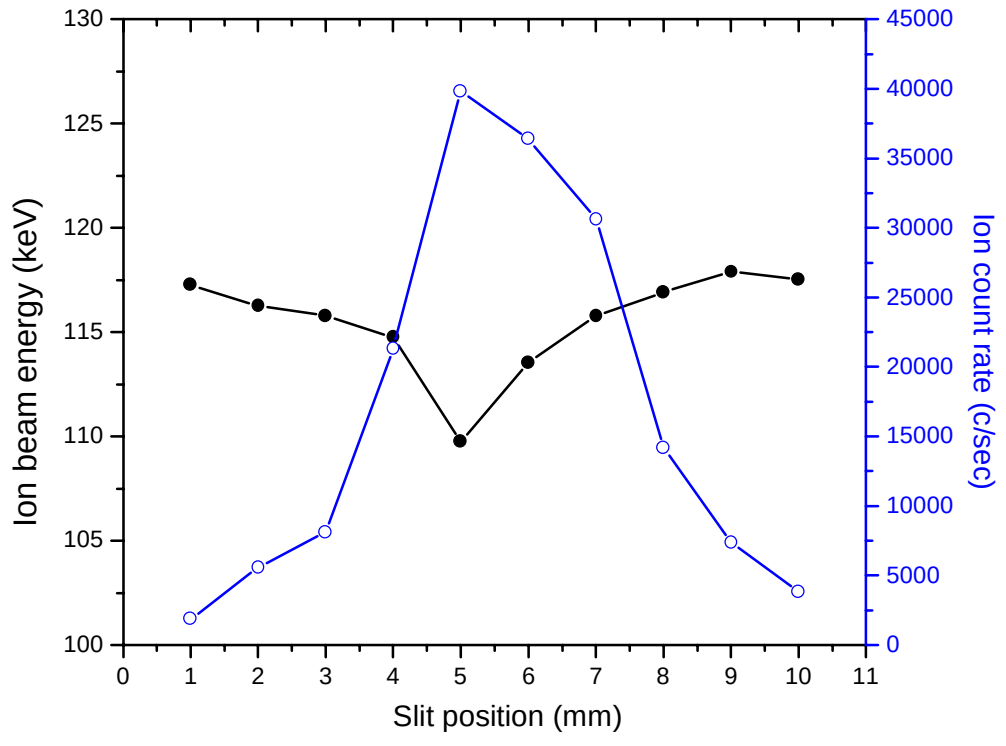

**Fig. 10**. The $N_2$+ ion beam count rate profile (open circles) and the beam energy (solid circles) at each slit position. This should be compared with the electron beam results in Fig. 7, where the beam energy is unaffected by the slit being at the focal position.

Figure 10 shows the $N_2$+ ion beam profile, which is also about 4mm when the crystal-to-slit distance is 23 mm (the focal length). On the other hand the $N_2$+ ion beam energy decreases by about 5% as the slit passes through the focal point of the beam as indicated by the count rate increase.

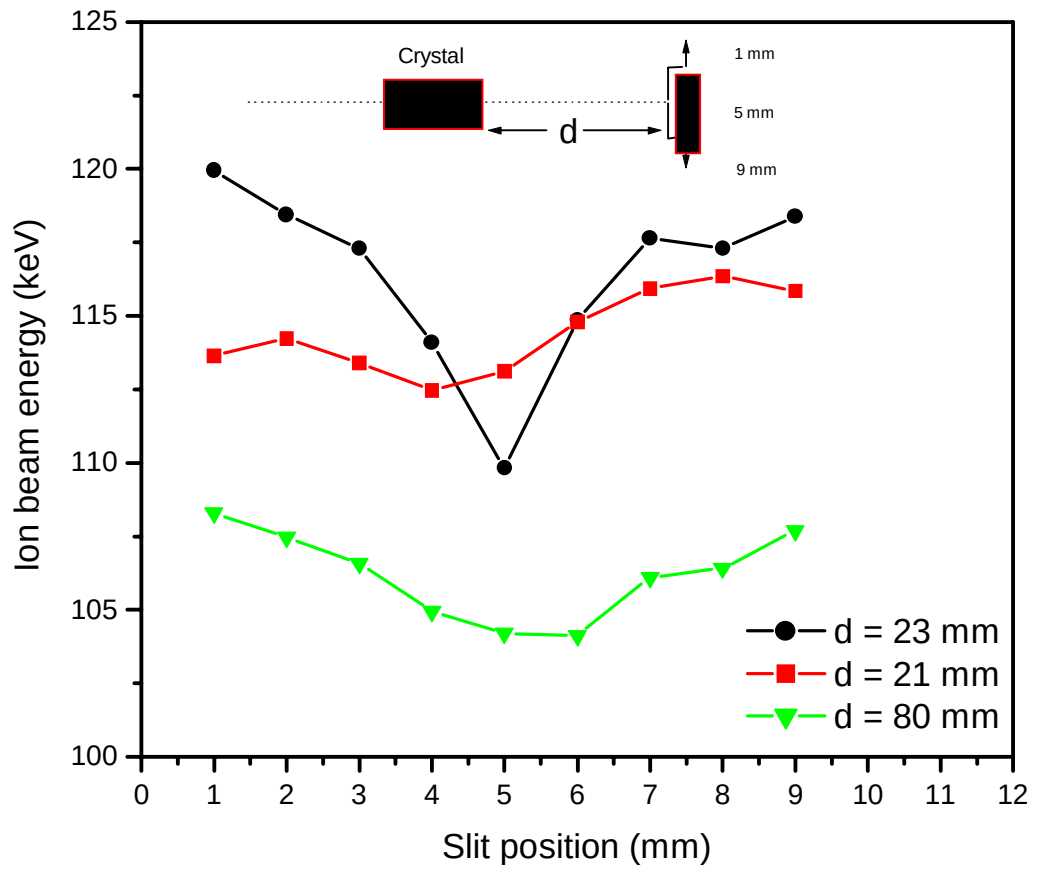

**Fig. 11.** Ion beam energies at different slit positions and at different crystal-to-slit distances. Note that the maximum ion beam energy decrease occurs at the crystal focal length, which is 23mm.

Finally Fig. 11 shows results of a study of the $N_2$+ ion beam energy vs slit position for three different crystal-to-slit distances. In this case the pressure-sensitive energy-drop at 23 mm (the focal length) is nearly 10%.

In conclusion self-focused energetic (<170 keV) electron beams are produced by heating to 160$^0$C cylindrical $LiNbO_3$ crystals with the - z base exposed in dilute gases at 2-8 mtorr on cooling. Conversely, positive gas-ion beams are produced by heating $LiNbO_3$ crystals when the + z base is exposed again on cooling.

## Acknowledgments

We are grateful to our colleagues, Tom Clegg, Bill Hooke, Sol Raboy, Brian Stoner and Eugen Merzbucher for insightful discussions.

[a)] e-mail jdbjdb@binghamton.edu
[b)]e-mail shafroth@physics.unc.edu